\documentclass{PoS}
 
\usepackage{graphicx}
\usepackage{multicol}

\newcommand{\treeC}[1] {\left[#1\right]_{C_{i}}} 
\newcommand{\be}{\begin{equation}}
\newcommand{\ee}{\end{equation}}
\newcommand{\ba}{\begin{eqnarray}}
\newcommand{\ea}{\end{eqnarray}}

\title{Determination of Low Energy Constants and testing Chiral Perturbation
  Theory at order {\boldmath $p^6$} (NNLO)}
\ShortTitle{Determination of LECs and testing ChPT at order $p^6$ (NNLO)}

\author{Johan Bijnens\\
        Department of Theoretical Physics, Lund University,
        S\"olvegatan 14A, SE 22362 Lund, Sweden\\
        E-mail: \email{bijnens@thep.lu.se}}

\author{\speaker{Ilaria Jemos}\\
        Department of Theoretical Physics, Lund University,
        S\"olvegatan 14A, SE 22362 Lund, Sweden\\
        E-mail: \email{ilaria.jemos@thep.lu.se}}

\abstract{We present the results of a search for relations between
observables that are independent of the Chiral Perturbation Theory (ChPT)
Next-to-Next-to-Leading Order (NNLO) Low-Energy Constants (LECs). We have found
some relations between observables in $\pi\pi$, $\pi K$ scattering and
$K_{l4}$ decay which have been evaluated numerically using fit 10
in~\cite{Amoros:2001cp} for the
NLO LECs. We also show some preliminary results for a new global fit of the NLO LECs}

\FullConference{6th International Workshop on Chiral Dynamics\\
                 July 6-10 2009\\
                 Bern, Switzerland}
\renewcommand{\logo}
{\raisebox{-5mm}
{\parbox{\textwidth}{\begin{flushright}
LU TP 09-24\\
September 2009
\end{flushright}
\includegraphics{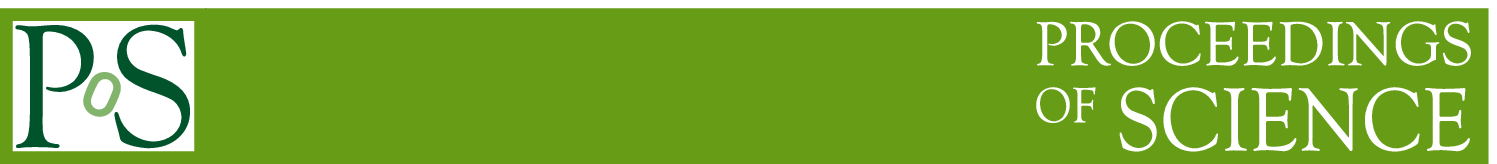}}
}}
\FullConference{{\rm To be published in the proceedings of:\\}
6th International Workshop on Chiral Dynamics\\
                 July 6-10 2009\\
                 Bern, Switzerland}

\begin{document}
\section{Introduction}
Testing the validity of Chiral Perturbation Theory (ChPT) is a challenging
task because of the many unknown parameters, called the Low Energy Constants
(LECs), entering into the theory. In particular, at NNLO $90$ unknown
constants, the $C_{i}$, appear in the $p^{6}$ Lagrangian. 

One way to overcome this problem is to study different combinations of
observables that depend on the $C_{i}$ in the same way. These lead to
$C_{i}$-independent relations which can be used to perform the
test. Furthermore 
those combinations might be useful to gain information on the LECs
too, since they let us isolate the same combinations of $C_{i}$ using
different observables.

In~\cite{Bijnens:2009zd}
 we study $76$ observables at NNLO and find $36$ such relations. We
 compare ChPT NNLO predictions with data/dispersive results for $13$ of
 these. The observables involved are the ones in $\pi
 \pi$ and $\pi K$-scattering and in $K_{\ell 4}$ decay.
Here we first discuss how we perform the numerical analysis, the results of
which appear
in Tab.\ref{tabpipi}, \ref{tabpipi2}, \ref{tabpiK} and \ref{tabpipipiK},
then we present for each process the relations studied. Finally we show some preliminary results for a new global fit of the $L_{i}$ at NNLO.
\section{Numerical Analysis}
The numerical analysis of the $C_{i}$-independent relations has been done in
the following way. First we evaluate the combinations of observables appearing in each side of
the relations using experiment/dispersive (exp) results of~\cite{CGL} for $\pi \pi$
scattering,~\cite{BDM} for $\pi K$ scattering
and~\cite{Batley:2007zz,Pislak:2001bf} for $K_{\ell 4}$ decay. Then we
 use ChPT results up to order $p^{6}$ \cite{BDT1,BDT,Amoros:2000mc} setting the $L_i$ to the
 values of fit 10 in~\cite{Amoros:2001cp}. Finally we subtract from the first (exp)
 evaluation the ChPT one. These differences will contain the $C_{i}$ part
 and higher order corrections. They have been quoted in
 Tab.\ref{tabpipi}, \ref{tabpipi2}, \ref{tabpiK} and \ref{tabpipipiK} in the
 columns labeled remainder. This has been done for each side of the relations under study. To check whether a relation is well satisfied we compare the remainders of its left-hand-side (LHS) and right-hand-side (RHS). Since they contain the same $C_{i}$ combinations, they should be equal within the uncertainties.

The errors quoted in the second columns of Tab.\ref{tabpipi}, \ref{tabpipi2}, \ref{tabpiK} and \ref{tabpipipiK}
are obtained adding in quadrature the uncertainties in
~\cite{CGL,BDM,Batley:2007zz,Pislak:2001bf}. This
might result in an underestimate of the total error because of
correlations. The theoretical errors due to the NLO LECs are shown in brackets
in the columns of Tab.\ref{tabpipi}, \ref{tabpipi2}, \ref{tabpiK} and \ref{tabpipipiK}  labeled NNLO 1-loop. They are obtained by varying all the $L_i$ around the central
values of fit 10 according to the full covariance matrix as obtained by the authors
of~\cite{Amoros:2001cp} and exploring the region with
$\chi^2/\textrm{dof}\approx 1$ . The error is then estimated as the
maximum deviation observed. The error for the $L_i$
contribution at NLO is never shown since it drops out of all the relations. No uncertainties due to higher order contributions have been added. The uncertainties due to theoretical errors are mostly on the last quoted digit.
\section{$\pi\pi$ scattering}

The $\pi\pi$ scattering amplitude can be written as a function $A(s,t,u)$
which is symmetric in $t,\,u$:
\begin{equation}
A(\pi^{a}\pi^{b}\rightarrow\pi^{c}\pi^{d})=
\delta^{a,b}\delta^{c,d}A(s,t,u)+\delta^{c,d}\delta^{b,d}A(t,u,s)
+\delta^{a,d}\delta^{b,c}A(u,t,s)\,,
\end{equation}
where $s,t,u$ are the usual Mandelstam variables.
 The isospin amplitudes $T^{I}(s,t)$ $(I=0,1,2)$ are
$
T^{0}(s,t)=3A(s,t,u)+A(t,u,s)+A(u,s,t)\,,\,
T^{1}(s,t)=A(s,t,u)-A(u,s,t)\,,\,
T^{2}(s,t)=A(t,u,s)+A(u,s,t)\,,\,
$
and are expanded in partial waves 
\ba
T^{I}(s,t)&=&32\pi\sum_{
  \ell=0}^{+\infty}(2\ell+1)P_{\ell}(\cos{\theta})t_{\ell}^{I}(s),
\ea
where $t$ and $u$ have been written as $t=-\frac{1}{2}(s-4m^{2}_{\pi})(1-\cos{\theta})$,
 $u=-\frac{1}{2}(s-4m^{2}_{\pi})(1+\cos{\theta})$.
Near threshold the $t^I_{\ell}$ are further expanded in terms of the threshold parameters
\begin{eqnarray}
t_{\ell}^{I}(s)=q^{2\ell}( a_{\ell}^{I}+ b_{\ell}^{I}q^{2}+\mathcal{O}(q^{4})),\qquad
 q^{2}=\frac{1}{4}(s-4m^{2}_{\pi}) ,
\end{eqnarray}
where $a_{\ell}^{I}, b_{ \ell}^{I}\dots$ are the 
scattering lengths, slopes,$\dots$.
We studied the $11$ parameters where a dependence on the $C_{i}$ shows
up.
Using $s+t+u = 4 m_\pi^2$ we can write the amplitude to order $p^6$ as
\begin{eqnarray}
\label{Apipi}
A(s,t,u)=b_{1}+b_{2}s+b_{3}s^{2}+b_{4}(t-u)^{2}+b_{5}s^{3}+b_{6}s(t-u)^{2}
+\textrm{non polynomial part}
\end{eqnarray}
The tree level Feynman diagrams give polynomial contributions to
$A(s,t,u)$ which must be expressible in terms of $b_1,\dots,b_6$.
Therefore we expect and find $5$ relations:
\ba
\label{pipi1}
 \treeC{5b^2_0- 2b^0_0 
  - 27 a^1_1 - 15a^2_0 + 6a^0_0}&=& -  18\treeC{b^1_1}
,\\
\label{pipi2}
\treeC{3a^1_1+b^2_0} &=&
 20\treeC{b^2_2 - b^0_2- a^2_2+a^{0}_2},\\
\label{pipi3}
\treeC{b^0_0+5b^2_0 +9a^1_1}
& = &
90\treeC{a^0_2-b^0_2}
,\\
\label{pipi4}
       \treeC{3b^{1}_1 + 25a^{2}_2} &=& 10\treeC{a^0_2},\\
\label{pipi5}
  \treeC{- 5b^2_2 + 2b^0_2} &=& 21\treeC{a^1_3},
\ea
where
$\treeC{A} \equiv C_i^r\textrm{-dependent~part~of~}A$.
All quantities are expressed in units of $m_{\pi^+}^2$.
In fact, since these relations hold for every
contribution to the polynomial part, they are valid for the NLO tree level
contribution as well and for two- and three-flavour ChPT.
Thus they get $L_i$-contributions only at NNLO via the non polynomial part of Eq.~(\ref{Apipi}).

In Tab.~\ref{tabpipi} we show our numerical results. We quote the left-hand-side (LHS) and right-hand-side
 (RHS) of each of
the relations. In the second column we use the values of the threshold parameters of
\cite{CGL}. 
The next columns use the ChPT results of~\cite{BDT1} and give the contributions from pure one-loop at NLO,
 the tree level NLO contribution,
 the pure two-loop contribution,
and the $L_i$ dependent part at NNLO (called NNLO 1-loop).

Comparing the remainders of the LHS with the RHS ones, we see that  the first three relations are very well satisfied, while the last two work at a level around two sigma. 

\begin{table}
\centering
\begin{tabular}{|c|c|r|r|r|r|c|}
\hline
                  & \cite{CGL}            & NLO    & NLO    &NNLO    &NNLO    & remainder\\
                  &                       & 1-loop &LECs    &2-loop  &1-loop  &          \\
\hline
LHS (\ref{pipi1})       & $ 0.009\pm0.039$&$ 0.054$&$-0.044$&$-0.041$&$-0.002(3) $&$ 0.041\pm0.039$\\
RHS (\ref{pipi1})       & $-0.102\pm0.002$&$-0.009$&$-0.044$&$-0.060$&$-0.008(6) $&$ 0.018\pm0.002$\\
10 LHS (\ref{pipi2})    & $ 0.334\pm0.019$&$ 0.209$&$ 0.097$&$ 0.103$&$ 0.029(11)$&$-0.105\pm0.019$\\
10 RHS (\ref{pipi2})    & $ 0.322\pm0.008$&$ 0.177$&$ 0.097$&$ 0.120$&$ 0.034(13)$&$-0.107\pm0.008$\\
LHS (\ref{pipi3})       & $ 0.216\pm0.010$&$ 0.166$&$ 0.029$&$ 0.053$&$ 0.016(6)$&$-0.047\pm0.010$\\
RHS (\ref{pipi3})       & $ 0.189\pm0.003$&$ 0.145$&$ 0.029$&$ 0.049$&$ 0.020(7)$&$-0.054\pm0.003$\\
10 LHS (\ref{pipi4})    & $ 0.213\pm0.005$&$ 0.137$&$ 0.032$&$ 0.053$&$ 0.035(12)$&$-0.043\pm0.005$\\
10 RHS (\ref{pipi4})    & $ 0.175\pm0.003$&$ 0.121$&$ 0.032$&$ 0.050$&$ 0.029(10)$&$-0.057\pm0.003$\\
$10^3$ LHS (\ref{pipi5})& $ 0.92\pm0.07  $&$ 0.36$ &$ 0.00$ &$ 0.56$ &$-0.01(13)$ &$0.00\pm0.07 $\\
$10^3$ RHS (\ref{pipi5})& $ 1.18\pm0.04  $&$ 0.42$ &$ 0.00$ &$ 0.57$ &$ 0.03(13)$ &$0.15\pm0.04 $\\
\hline
\end{tabular}
\caption{\label{tabpipi} The relations found in the $\pi\pi$-scattering.
The lowest order contribution is always zero by construction.
The NLO LEC part satisfies the relation, as it should. Notice the extra factors of
ten for some of them. All quantities are in the units of powers of $m_{\pi^+}$.
}
\end{table}

We can also check how the two-flavour predictions hold up.
Since here the corrections are in powers
of $m_\pi^2$ rather than in powers of  $m_K^2$, the expansion should 
converge better. For the ChPT evaluation we use the threshold parameters as
quoted in \cite{CGL} for their best fit of the NLO LECs. The result is shown in Tab.~\ref{tabpipi2}.
We see the same pattern as for the three-flavour case: the first three relations are very well satisfied while the last two
are somewhat worse but below two sigma.
\begin{table}
\centering
\begin{tabular}{|c|c|c|c|}
\hline
                  & \cite{CGL}            & two-flavour & remainder\\
                  &                       & \cite{CGL} &          \\
\hline
LHS (\ref{pipi1})       & $ 0.009\pm0.039$&$-0.003$&$ 0.007\pm0.039$\\
RHS (\ref{pipi1})       & $-0.102\pm0.002$&$-0.097$&$-0.005\pm0.002$\\
10 LHS (\ref{pipi2})    & $ 0.334\pm0.019$&$0.332$&$ 0.002\pm0.019$\\
10 RHS (\ref{pipi2})    & $ 0.322\pm0.008$&$0.318$&$ 0.004\pm0.075$\\
LHS (\ref{pipi3})       & $ 0.216\pm0.010$&$0.206$&$ 0.010\pm0.010$\\
RHS (\ref{pipi3})       & $ 0.189\pm0.003$&$0.189$&$ 0.000\pm0.003$\\
10 LHS (\ref{pipi4})    & $ 0.213\pm0.005$&$0.204$&$ 0.009\pm0.005$\\
10 RHS (\ref{pipi4})    & $ 0.175\pm0.003$&$0.176$&$-0.001\pm0.003$\\
$10^3$ LHS (\ref{pipi5})& $ 0.92\pm0.07  $&$1.00$ &$-0.08 \pm0.07 $\\
$10^3$ RHS (\ref{pipi5})& $ 1.18\pm0.04  $&$1.15$ &$ 0.04 \pm0.04 $\\
\hline
\end{tabular}
\caption{\label{tabpipi2} The relations found in the $\pi\pi$-scattering
  evaluated in two-flavour ChPT. In the second column we have used the
 NNLO results
  quoted in \cite{CGL}. Notice the extra factors of
ten for some of them. All quantities are in units of powers of $m_{\pi^+}$.
}
\end{table}

\section{$\pi K$ scattering}

The $\pi K$ scattering has amplitudes $T^I(s,t,u)$
in the isospin channels $I= 1/2,3/2$.
 As for $\pi \pi$ scattering we introduce the partial wave expansion of
 the isospin amplitudes
\begin{eqnarray}
T^{I}(s,t,u)&=&16\pi
\sum_{\ell=0}^{+\infty}(2\ell+1)P_{\ell}(\cos{\theta})t_{\ell}^{I}(s),
\end{eqnarray}
and we define scattering lengths
 $a_{\ell}^{I}$, $b_{\ell}^{I}$ by expanding the $t_{\ell}^{I}(s)$ near threshold:
\begin{eqnarray}
t_{\ell}^{I}(s)&=&\frac{1}{2}\sqrt{s}q_{\pi K}^{2\ell}
\left(a_{\ell}^{I}+b^{I}_{\ell}q^{2}_{\pi K}+\mathcal{O}(q^{4}_{\pi K})\right),
\qquad
q_{\pi K}^{2}=\frac{s}{4}\left(1-\frac{(m_{K}+m_{\pi})^{2}}{s}\right)
          \left(1-\frac{(m_{K}-m_{\pi})^{2}}{s}\right)\,,
\nonumber
\end{eqnarray}
and
$t=-2q^{2}_{\pi K}(1-\cos{\theta}),\quad u=-s-t+2m^{2}_{K}+2m^{2}_{\pi}$.
Again we studied only those observables where a
dependence on the $C_{i}$ shows up.
 
It is also customary to introduce the crossing symmetric and antisymmetric
amplitudes $T^{\pm}(s,t,u)$ 
\begin{eqnarray}
3T^{+}(s,t,u)&=&T^{1/2}(s,t,u)+T^{3/2}(s,t,u), \qquad
T^{-}(s,t,u)=T^{1/2}(s,t,u)-T^{3/2}(s,t,u),
\end{eqnarray}
 which can be expanded around
$t=0$, $s=u$ using $\nu= (s-u)/(4m_{K})$ (subthreshold expansion):
\ba
\label{subthrexp}
T^{+}(s,t,u)&=&\sum_{i,j=0}^{\infty}c_{ij}^{+}t^{i}\nu^{2j},\qquad
T^{-}(s,t,u)=\sum_{i,j=0}^{\infty}c_{ij}^{-}t^{i}\nu^{2j+1}.
\ea
There are 10 subthreshold parameters that have tree level contributions
 from the NNLO LECs.
In $c_{01}^{-}$ and $c_{20}^{-}$ the same combination
$-C_{1}+2C_{3}+2C_{4}$ appears~\cite{BDT}:
\be
\label{piKrel}
16\rho^{2}\treeC{c_{20}^{-}}=3\treeC{c_{01}^{-}}\,.
\ee
Therefore in the isospin odd channel only three subthreshold parameters get independent
 contributions from
the $C_{i}$. So for the 7 differences $a^-_\ell = a_\ell^{1/2}-a^{3/2}_\ell$
and $b^-_\ell = b_\ell^{1/2}-b^{3/2}_\ell$ getting contributions at NNLO and three
subthreshold parameters
we expect four relations:
\ba
\label{piK1}
&&\left(\rho^4+3\rho^3+3\rho+1\right)\treeC{a_1^-}=
2\rho^2\left(\rho+1\right)^2\treeC{b_1^-}
-\frac{2}{3}\rho\left(\rho^2+1\right)\treeC{b_0^-}
\nonumber\\&&
+\frac{1}{2\rho}\left(\rho^2+\frac{4}{3}\rho+1\right)\left(\rho^2+1\right)
\treeC{a_0^-}\,,
\\
\label{piK3}
&&5\left(\rho+1\right)^2\treeC{b_2^-}
=\frac{\left(\rho-1\right)^2}{\rho^2}\treeC{a_1^-}
-\frac{\rho^4+\frac{2}{3}\rho^2+1}{4\rho^4}\treeC{a_0^-}
+\frac{\rho^2-\frac{2}{3}\rho+1}{2\rho^2}\treeC{b_0^-}\,,\\
\label{piK2}
&& 5\left(\rho^2+1\right)\treeC{a_2^-}
 =
\treeC{a_1^-}
+2\rho\treeC{b_1^-}\,,
\\
\label{piK4}
&& 7\left(\rho^2+1\right)\treeC{a_3^-}
 =
\treeC{a_2^-}
+2\rho\treeC{b_2^-}\,,
\ea
the threshold parameters are expressed in units of $m_{\pi^+}$ and we use
the symbol $\rho=m_K/m_\pi$.

$T^+$ brings in 7 more combinations of threshold parameters, $a^+_\ell = a_\ell^{1/2}+2a^{3/2}_\ell$
and $b^+_\ell = b_\ell^{1/2}+2b^{3/2}_\ell$, but there are 6
 independent subthreshold
parameters so we find only one more relation:
\ba
\label{piK5}
7\treeC{a_3^+} &=&\frac{1}{2\rho}\treeC{a_2^+}
-\treeC{b_2^+}
+\frac{1}{5\rho}\treeC{b_1^+}
-\frac{1}{60\rho^3}\treeC{a_0^+}
-\frac{1}{30\rho^2}\treeC{b_0^+}\,.
\ea

Again these relations hold for all tree-level contributions up to 
NNLO.
\begin{table}
\centering
\begin{tabular}{|c|c|r|r|r|r|c|}
\hline
                  & \cite{BDM}         & NLO   & NLO  &NNLO    &NNLO    & remainder\\
                  &                    & 1-loop&LECs  &2-loop  &1-loop  &          \\
\hline
LHS (\ref{piK1})       & $ 5.4\pm0.3  $&$0.16$&$ 0.97$&$ 0.77$&$-0.11(11)$&$ 0.6\pm0.3$\\
RHS (\ref{piK1})       & $ 6.9\pm0.6  $&$0.42$&$ 0.97$&$ 0.77$&$-0.03(7)$&$ 1.8\pm0.6$\\
10 LHS (\ref{piK2})    & $ 0.32\pm0.01$&$0.03$&$ 0.12$&$ 0.11$&$ 0.00(2)$&$ 0.07\pm0.01$\\
10 RHS (\ref{piK2})    & $ 0.37\pm0.01$&$0.02$&$ 0.12$&$ 0.10$&$-0.01(2)$&$ 0.14\pm0.01$\\
100 LHS (\ref{piK3})   & $-0.49\pm0.02$&$0.08$&$-0.25$&$-0.17$&$ 0.05(3)$&$-0.21\pm0.02$\\
100 RHS (\ref{piK3})   & $-0.85\pm0.60$&$0.03$&$-0.25$&$ 0.11$&$-0.03(13)$&$-0.71\pm0.60$\\
100 LHS (\ref{piK4})   & $ 0.13\pm0.01$&$0.04$&$ 0.00$&$ 0.01$&$ 0.03(1)$&$ 0.05\pm0.01$\\
100 RHS (\ref{piK4})   & $ 0.01\pm0.01$&$0.01$&$ 0.00$&$ 0.00$&$ 0.00(1)$&$-0.01\pm0.01$\\
$10^3$ LHS (\ref{piK5})& $ 0.29\pm0.03$&$0.09$&$ 0.00$&$ 0.06$&$ 0.01(2)$&$ 0.13\pm0.03$\\
$10^3$ RHS (\ref{piK5})& $ 0.31\pm0.07$&$0.03$&$ 0.00$&$ 0.06$&$ 0.05(3)$&$ 0.17\pm0.07$\\
\hline
\end{tabular}
\caption{\label{tabpiK} The relations found in the $\pi K$-scattering.
The tree level contribution to the LHS and RHS
 of relation 1
is 3.01 and vanishes for the others.
The NLO LECs part satisfies the relation. Notice the extra factors of
ten for some of them. All quantities are in the units of powers of $m_{\pi^+}$
}
\end{table}
The numerical check is shown in Tab.~\ref{tabpiK}.
The columns in Tab.~\ref{tabpiK} have the same meaning as in
 Tab.~\ref{tabpipi}.

The first relation is reasonably satisfied, somewhat below two sigma.
 The second
relation has a large discrepancy but
if we assume
a theory error of about half the NNLO contribution it seems reasonable.
 The third relation
is well satisfied but the RHS has a rather large experimental error.
The fourth relation does not work well, mainly due to the fact that we seem to
underestimate the value for $a^-_3$. The last relation works well.
 
\section{$\pi\pi$ and $\pi K$ scattering}
If we consider the $\pi\pi$ and $\pi K$ system
 together we get two more relations due to the identities
\ba
\treeC{b_{5}}&=&\treeC{c^{+}_{30}}+\frac{3}{\rho}\treeC{c^{-}_{20}}\,,
\qquad
\treeC{b_{6}}=\frac{1}{4\rho}\treeC{c^{-}_{20}}
   +\frac{1}{16\rho^2}\treeC{c^{+}_{11}},
\ea
where $c^{-}_{ij}$ ($c^{+}_{ij}$) are  expressed
in units of $m^{2i+2j+1}_{\pi}$($m^{2i+2j}_{\pi}$). 
We can express these relations in terms of the threshold parameters (all quantities expressed in powers of $m_{\pi^+}$):
\ba
\label{pipipiK1}
&&6 \treeC{a^1_3}=\left(1+\rho\right)\treeC{a^+_3+3 a^-_3}\,,
\\
\label{pipipiK2}
&&3\left[\left(1+\rho\right)^2 \treeC{b^2_2}
 \hspace{-0.15cm}+ 7\left(1-\rho\right)^2 \treeC{a^1_3} \right]=
\left(1+\rho\right)\left[7 \left(1-4\rho+\rho^2\right)\treeC{a^-_3}
+\treeC{a^+_2+2\rho b^+_2}\right].\,\,\qquad
\ea
The numerical results are quoted in Tab.~\ref{tabpipipiK}. The first
 relation does not work
but the second is well satisfied. If we look in the numerical results
 we see that $a^-_3$
plays a minor role in the RHS of the second relation but is important
 in the first, so this
could be the same problem of relation (\ref{piK4}).
A related analysis can be found in \cite{KM}.

\begin{table}
\centering
\begin{tabular}{|c|c|r|r|r|r|c|}
\hline
                  & \cite{CGL},\cite{BDM}  & NLO   & NLO  &NNLO    &NNLO    & remainder\\
           & \cite{Batley:2007zz},\cite{Pislak:2001bf} & 1-loop&LECs  &2-loop  &1-loop  &          \\
\hline
$10^3$ LHS (\ref{pipipiK1})& $0.34\pm0.01$&$ 0.12$&$0.00$&$ 0.16$&$ 0.00(4)$&$ 0.05\pm0.01$\\
$10^3$ RHS (\ref{pipipiK1})& $0.38\pm0.03$&$ 0.12$&$0.00$&$ 0.05$&$ 0.04(2)$&$ 0.16\pm0.03$\\
10 LHS (\ref{pipipiK2})   & $-0.13\pm0.01$&$-0.12$&$0.00$&$-0.05$&$ 0.02(2)$&$ 0.01\pm0.01$\\
10 RHS (\ref{pipipiK2})   & $-0.09\pm0.02$&$-0.05$&$0.00$&$-0.02$&$-0.01(1)$&$-0.01\pm0.02$\\
\hline
LHS (\ref{Kl4rel}) & $-0.73\pm0.10$ & $-0.23$ & $0.00$ & $-0.15$ & $-0.05(6)$ & $-0.29\pm0.10$\\ 
RHS (\ref{Kl4rel}) &  $0.50\pm0.07$ &  $0.19$ & $0.00$ &  $0.10$ &  $0.03(4)$ & $0.18\pm0.07$\\
\hline
\end{tabular}
\caption{\label{tabpipipiK} The relations found between $\pi\pi$ and $\pi
  K$-scattering lengths and between the curvature in $F$ in $K_{\ell4}$
 and $\pi K$ scattering. All quantities are in the units of powers of $m_{\pi^+}$.
}
\end{table}

\section{\boldmath$K_{\ell4}$}
\label{Kl4}

The decay $K^+(p)\to \pi^+(p_1)\pi^-(p_2) e^+(p_\ell)\nu(p_\nu)$ is given 
by the amplitude~\cite{Bijnens:1994me}
\be
      T = \frac{G_F}{\sqrt{2}} V^\star_{us} \bar{u} (p_\nu) \gamma_\mu
      (1-\gamma_5) v(p_\ell) (V^\mu - A^\mu)
      \label{k11}
\ee
where $V^\mu$ and $A^\mu$ are parametrized in terms of four formfactors: $F$,
$G$, $H$ and $R$  (but the $R$-formfactor is negligible in decays with
an electron in the final state).
Using partial wave expansion and neglecting $d$ wave terms one
obtains~\cite{Amoros:1999mg}:
\ba
\label{Kl4exp}
F&=&f_{s}+f'_{s}q^{2}+f''_{s}q^{4}+f'_{e}s_{e}/4m^{2}_{\pi}
+f_{t}\sigma_\pi X\cos\theta+\ldots
\,,
\nonumber\\
G_{p}&=&g_{p}+g'_{p}q^{2}+g''_{g}q^{4}+g'_{e}s_{e}/4m^{2}_{\pi}
+g_{t}\sigma_\pi X\cos\theta+\ldots
\ea
Here
$s_{\pi}(s_{e})$ is the invariant mass of dipion (dilepton) system, and
$ q^{2}=s_{\pi}/(4m^{2}_{\pi})-1$.
$\theta$ is the angle of the pion in their rest frame
w.r.t. the kaon momentum and $t-u=-2\sigma_\pi X\cos\theta$.  
Using NNLO ChPT results~\cite{BDT,Amoros:2000mc} we find one relation between the quantities defined in (\ref{Kl4exp}) and
$\pi K$ scattering:
\be
\label{Kl4r}
\sqrt{2}\treeC{f''_{s}}= 64 \rho F_\pi\treeC{c^{+}_{30}}\,.
\ee
This leads to a relation between $\pi K$ threshold parameters
and $f''_{s}$ which, with all quantities expressed in units of $m_{\pi^+}$,
reads:
\be
\label{Kl4rel}
\sqrt{2}\treeC{f''_{s}}= 32\pi\frac{\rho}{1+\rho} F_\pi
\left[\frac{35}{6}\left(2+\rho+2\rho^2\right)\treeC{a^+_3}
-\frac{5}{4}\treeC{a^+_2+2\rho b^+_2}\right]\,.
\ee

Numerical results for (\ref{Kl4rel}) are shown in Tab.~\ref{tabpipipiK}.
The experimental results is taken from \cite{Batley:2007zz} for $f''_s/f_s$ and from
 \cite{Pislak:2001bf}
for $f_s$. This should be an acceptable combination since the central value
 for $f'_s/f_s$ and
$f''_s/f_s$ from \cite{Pislak:2001bf} are in good agreement with those of \cite{Batley:2007zz}.
This relation is not satisfied: the sign is even different on the two sides.
 Notice that, in both cases, we also
see that the ChPT series has a large NNLO contribution.

\begin{figure}
\begin{center}
\includegraphics[angle=270,width=10cm]{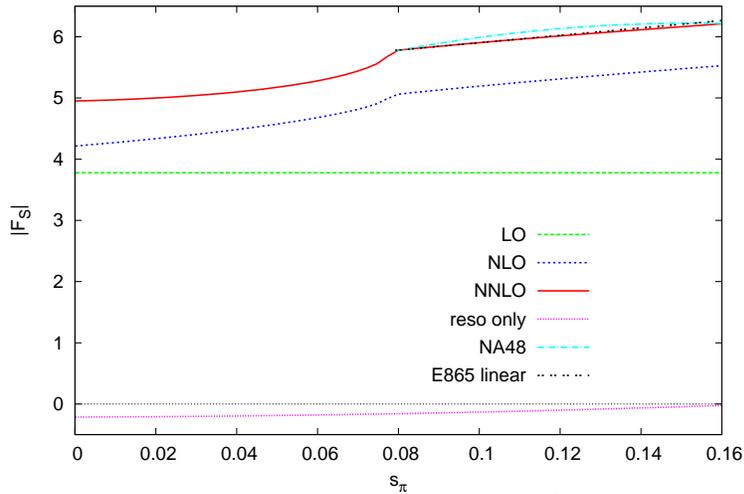}
\end{center}
\vspace{-1cm}
\caption{\label{kl4f_fit10} The absolute value of the $F_s$ formfactor at
  $s_\ell=\cos{\theta}=0$ as a function of $s_\pi$ (in
  $\textrm{Gev}^2$ units) above and below threshold.
  The NNLO result nicely reproduces the linear fit quoted in \cite{Pislak:2001bf}, but not the
  large negative curvature in \cite{Batley:2007zz}. The line at the bottom is the
  contribution coming from the $C_i$, which has a positive curvature.}
\end{figure}

It has been already noticed, see~\cite{Amoros:2001cp} and
Fig.~\ref{kl4f_fit10}, that ChPT, at present, underestimates the curvature $f''_s$.
On the other hand there are indications that dispersive
analysis techniques might help solving this problem: Fig.~7 in \cite{Amoros:2001cp} shows that the dispersive result
of \cite{BCG} has a larger curvature then the two-loop result.
Therefore, we do not consider this discrepancy a major problem for ChPT.
\vspace{-0.5cm} 
\section{New fits of the NLO constants (preliminary results)}
As remarked in \cite{Bijnens:2006zp}, many NNLO calculations are now available in
three-flavour ChPT. Besides, new lattice and dispersive results and
further experimental data are at our disposal too. A study of the predictive power
of NNLO ChPT is needed, and therefore also an update of the $L_i$ fit.\\
For this reason we are working on a new program to perform this fit with many
more observables implemented. So far we have included masses and decay
constants, $K_{\ell
  4}$ formfactors, $\pi \pi$
and $\pi K$ scattering lengths and the scalar pion radius. For now we rely on the resonance estimates of the $C_i$ used
in~\cite{Amoros:2001cp}, although our plan is to achieve more information on them.
\begin{table}
\centering
\begin{tabular}{cccccc}
             & fit 10 \cite{Amoros:2001cp} &fit 10 iso   & NA48/2 & $F_K/F_\pi$  &All       \\
 \hline
$10^3 L_1^r$ & $0.43$ &$0.40\pm0.12$ &  \framebox{$0.98$}  & $0.97$&$0.99\pm0.13$ \\
$10^3 L_2^r$ & $0.73$&$0.76\pm0.12$ & $0.78$   & $0.79$ &$0.60\pm0.22$   \\
$10^3 L_3^r$ &$ -2.35$&$-2.40\pm0.37$& \framebox{$-3.14$} &$-3.12$&$-3.07\pm0.59$ \\
$10^3 L_4^r$ &$\equiv0$ &$\equiv0$          & $\equiv0$& $\equiv0$  & \framebox{$0.65\pm0.64$}\\
$10^3 L_5^r$ &$0.97$ &$0.97\pm0.11$  & $0.93$     &  \framebox{ $0.72$}  & \framebox{$0.53\pm0.10$}   \\
$10^3 L_6^r$ &$\equiv0$ &$\equiv 0$         & $\equiv 0$& $\equiv 0$&  \framebox{$0.07\pm0.65$}\\
$10^3 L_7^r$ &$-0.31$ &$-0.30\pm0.15$& $-0.30$     &$-0.26$  &$-0.21\pm0.15$    \\
$10^3 L_8^r$ &$0.59$ &$0.61\pm0.20$ & $0.59$      & \framebox{$0.48$}    &$0.37\pm0.17$ \\
\hline
$\chi^2$ (dof)&      & $0.25$ (1) &$0.17$ (1)&$0.19$ (1) &$0.78$ (4) 
\end{tabular} 
\caption{\label{tabfit} Preliminary results for the fits. $L_9 \equiv
  0.59\times10^{-3}$ everywhere, as found from the vector pion radius in
  \cite{BT2}. See text for a
  longer discussion}.
\end{table}

Our first preliminary results are summarized in Tab.~\ref{tabfit}.
In the second column we quote fit 10 of
\cite{Amoros:2001cp}. This was found using the available linear fit for $K_{\ell
  4}$ of \cite{Pislak:2001bf}, $F_K/F_\pi = 1.22 $, the kaon and eta masses
with isospin breaking corrections included and setting $L_4\equiv L_6\equiv 0$. In the column labeled fit 10 iso we quote the fit we
find using the same input as fit 10 but without including isospin breaking. As you
see the two fits are in good agreement.
The column NA48/2 relies on the new experimental data from~\cite{Batley:2007zz}. We checked that
the fit does not change including the curvature $f''_s$. With this fit ChPT
predicts the value $f''_s=-0.90$ to be compared with the experimental one
$f''_s=-1.58\pm0.064$. Note that the fit in~\cite{Batley:2007zz} shows large correlations
between the slope and the curvature of the $F_s$ formfactor which have
not been taken into account yet. The values of $L_1$ and $L_3$ change drastically.
The third column shows the fit obtained changing the
ratio $F_K/F_\pi$ to $1.19$. This affects mainly $L_5$ and $L_8$.
The last column shows the fit obtained letting $L_4$ and $L_6$ free, and
adding $a^0_0$, $a^2_0$, $a^{1/2}_0$, $a^{3/2}_0$ and the scalar pion
radius. The value obtained for $L_4$ is larger then expected. Some more
comment can be found in~\cite{Bijnens:chdyn09}.
\section{Conclusions}
\label{conclusions}
We have performed a systematic search for relations between observables that allow a test of
ChPT at NNLO
order in a $C_i$-independent way. We studied in detail
 the relations for the $\pi\pi$,
$\pi K$ scattering and $K_{\ell4}$ since for these cases enough experimental
 and/or dispersion
theory results exist.

 The resulting picture is that ChPT at NNLO mostly works but there are troublesome cases. The $\pi\pi$ system alone works well.
The $\pi K$ system alone works satisfactorily but with some discrepancies. The same
can be said for the combinations of both systems. A common part
in these two cases is the presence
of $a^-_3$ . Comparing $\pi K$ scattering and $K_{\ell4}$ leads to a clear
contradiction which needs further investigation.
\section*{Acknowledgments}
IJ gratefully acknowledges an Early Stage Researcher position supported by the
EU-RTN Programme, Contract No. MRTN--CT-2006-035482, (FLAVIAnet).
This work is supported in part by the European Commission RTN network,
Contract MRTN-CT-2006-035482  (FLAVIAnet), 
European Community-Research Infrastructure
Integrating Activity
``Study of Strongly Interacting Matter'' (HadronPhysics2, Grant Agreement
n. 227431)
and the Swedish Research Council. We thank the organizers for a very pleasant
meeting.

\end{document}